\documentclass[12pt]{spieman}  
\usepackage{setspace}
\usepackage{tocloft}

\usepackage{amsmath,amssymb,amstext} 
\usepackage{graphicx}
\usepackage{caption}
\usepackage{subcaption}
\usepackage{multirow}
\usepackage{array}
\usepackage{float}  
\usepackage{xspace}  
\usepackage{enumitem}  
\usepackage{authblk}
\usepackage{bbm}  

\renewcommand{\vec}[1]{\mathbf{#1}}

\DeclareMathOperator*{\argmax}{arg\,max}

\title{Spatial probabilistic pulsatility model for enhancing photoplethysmographic imaging systems}

\author[a,b*]{Robert~Amelard}
\author[a]{David~A~Clausi}
\author[a,b]{Alexander~Wong}
\affil[a]{University of Waterloo, Department of Systems Design Engineering, 200 University Ave W, Waterloo, Canada, N2L 3G1}
\affil[b]{Schlegel-University of Waterloo Research Institute for Aging, 250 Laurelwood Dr, Waterloo, Canada, N2J 0E2}

\begin{document}
\maketitle

\begin{abstract}
Photolethysmographic imaging (PPGI) is a widefield non-contact biophotonic technology able to remotely monitor cardiovascular function over anatomical areas. Though spatial context can provide increased physiological insight, existing PPGI systems rely on coarse spatial averaging with no anatomical priors for assessing arterial pulsatility. Here, we developed a continuous probabilistic pulsatility model for importance-weighted blood pulse waveform extraction. Using a data-driven approach, the model was constructed using a 23~participant sample with large demographic variation (11/12 female/male, age 11--60 years, BMI 16.4--35.1 kg$\cdot$m$^{-2}$). Using time-synchronized ground-truth waveforms, spatial correlation priors were computed and projected into a co-aligned importance-weighted Cartesian space. A modified Parzen-Rosenblatt kernel density estimation method was used to compute the continuous resolution-agnostic probabilistic pulsatility model. The model identified locations that consistently exhibited pulsatility across the sample. Blood pulse waveform signals extracted with the model exhibited significantly stronger temporal correlation ($W=35,p<0.01$) and spectral SNR ($W=31,p<0.01$) compared to uniform spatial averaging. Heart rate estimation was in strong agreement with true heart rate ($r^2=0.9619$, error $(\mu,\sigma)=(0.52,1.69)$~bpm).
\end{abstract}

{\noindent \footnotesize\textbf{*}Robert Amelard,  \linkable{ramelard@uwaterloo.ca} }
\begin{spacing}{2}   



\section{Introduction}
\label{sect:intro}

Photoplethysmography (PPG) is a non-invasive optical technique for cardiovascular monitoring~\cite{allen2007}. In its simplest form, a PPG device comprises an illumination source (e.g., LED) and detector (e.g., photodiode). By monitoring the temporal illumination changes, PPG devices measure the pulsatile blood pulse waveform from transient local arterial volume fluctuations. This information can be leveraged to monitor cardiovascular characteristics such as heart rate, heart rate variability, blood pressure, and cardiac output~\cite{allen2007}. However, existing contact-based systems are limited to single-location monitoring (e.g., finger), can only be used by one individual per device, and motion artifacts inhibit its use in ambulatory scenarios.

Photoplethysmographic imaging (PPGI) systems are biophotonic systems that have recently gained interest for non-contact widefield cardiovascular monitoring of cardiac parameters such as heart rate, breathing rate, pulse oxygen saturation, and heart rate variability~\cite{sun2016,allen2014}. PPGI systems extend PPG devices by decoupling the illumination source and detector from each other and the body. Most PPGI systems use a camera as the illumination detector, enabling widefield blood pulse imaging over large areas, enabling new types of monitoring such as spatial perfusion analysis~\cite{kamshilin2011,kamshilin2013} and multi-individual monitoring~\cite{poh2010}. However, many systems use coarse spatial averaging to estimate cardiovascular perfusion, such as averaging over the entire facial bounding box~\cite{poh2010,sun2011,sun2012} and hardcoded areas~\cite{kong2013,mcduff2014,li2014}. One study attained increased accuracy by incorporating spatial pulsatility priors~\cite{kumar2015}; however the method relies on a real-time estimate of the true physiological state based on aforementioned coarse averaging techniques to achieve accurate prediction.

In this paper, we developed a continuous probabilistic pulsatility model for importance-weighted blood pulse waveform extraction. The continuous model can be used by PPGI systems of any resolution through appropriate sampling to extract robust blood pulse waveforms. The model was developed using a data-driven approach over a 23~participant sample with highly varying characteristics (11/12~female/male, age 11--60~years, body fat 10.5--42.3\%, muscle 31.0--52.7\%, BMI 16.4--35.1~kg$\cdot$m$^{-2}$). Using blood pulse waveform spatial correlation priors, an importance weighting scheme was developed, giving locations with consistently strong pulsatility a higher weight in the final model. Samples were aligned and aggregated in a common Cartesian space, and the continuous probabilistic pulsatility model was computed using a kernel density estimation approach. This method was compared against whole-area uniform spatial averaging approach used by existing studies~\cite{poh2010,sun2011,sun2012}. Results showed that signals extracted using the pulsatility model were statistically significantly stronger in temporal (correlation, $p<0.01$) and spectral (SNR, $p<0.01$) characteristics than uniform averaging, and heart rates were in tight agreement with ground-truth measurements ($r^2=0.9619$, error $\mu=0.52\text{ bpm},\sigma=1.7\text{ bpm}$). Model visualization elucidated important arterial pathways, including the neck, malar regions, glabella regions, lips and nose. We discuss how the model can be used and trained for custom applications.

\section{Methods}
The goal was to compute a continuous probabilistic spatial pulsatility model that can be used as \textit{a priori} information in PPGI systems. By computing a continuous model, it can be used by datasets of any resolution through appropriate discrete spatial sampling. Figure~\ref{fig:pipeline} depicts the processing pipeline to generate this pulsatility model. This study was approved by a Research Ethics Committee at the University of Waterloo. Informed consent was obtained from all participants prior to data collection. Additionally, informed consent was obtained from those individuals whose photos were used in this paper.

\subsection{Data Collection}
A novel photoplethysmographic imaging system, coded hemodynamic imaging (CHI)~\cite{amelard2015vislet,amelard2015scirep}, was used to collect the imaging data. The scene was illuminated with a diffuse uniform broadband tungsten-halogen illumination source using a glass fabric front diffuser (Lowel Rifa eX 44). Images were acquired at 60~fps using a monochrome CMOS camera (Point Grey GS3-U3-41C6NIR-C) with NIR sensitivity. A 850--1000~nm bandpass filter was mounted in front of the lens to constrain the sensor measurements to deep NIR tissue penetration. Participants ($n=23$) were instructed to remain supine for the duration of the study. The camera was positioned overhead at 1.5~m from the participant's head. The camera angle and field of view remained fixed across participants. Participants wore a finger photoplethysmography cuff, providing a ground-truth blood pulse waveform signal synchronously with the video frames.

\begin{table}
\centering
\caption{Sample demographics}
\begin{tabular}{|lc|}
\hline
Demographic & Sample Representation \\
\hline
n (male/female) & 23 (12/11) \\
age (years) & 11 -- 60 $(29.6 \pm 11.9)^*$ \\
mass (kg) & 42.6 -- 107.0 $(73.4 \pm 17.8)^*$ \\
height (cm) & 145 -- 193 $(168.5 \pm 11.4)^*$ \\
body fat (\%) & 10.5 -- 42.3 $(21.3 \pm 7.8)^*$ \\
muscle (\%) & 31.0 -- 52.7 $(39.7 \pm 4.5)^*$ \\
BMI (kg$\cdot$m$^{-2}$) & 16.4 -- 35.1 $(25.7 \pm 5.2)^*$ \\
\hline
\multicolumn{2}{l}{$^*\mu \pm \sigma$}
\end{tabular}
\label{tab:demographics}
\end{table}

\subsection{Probabilistic pulsatility model}
Different spatial locations exhibit varying amounts of observed pulsatility. For example, a skin location directly above a shallow major artery will exhibit high pulsatility due transient changes in local tissue optical properties from the arterial pulse~\cite{allen2007,kamshilin2015}, whereas occluding hair will inhibit the observed pulsatility. Let $p(x,y)$ be the probabilistic pulsatility model such that $p(x,y)$ quantifies the probability of observing arterial pulsatility at location $(x,y)$. Using a physiologically-motivated data-driven approach, $p(x,y)$ can be computed by determining the locations that consistently exhibited pulsatility across a diverse sample of participants (see Table~\ref{tab:demographics}). Such a model can be used to as an \textit{a priori} model for new data when extracting cardiovascular properties, such as heart rate or the blood pulse waveform.

\subsubsection{Absorbance mapping}
Let $f(x,y,t)$ be a set of frames. The blood pulse waveform signal is typically represented by transient changes in absorbance rather than reflectance, thus $f$ was transformed to absorbance:
\begin{equation}
a(x,y,t) = -\log(f(x,y,t))
\label{eq:a}
\end{equation}
Each signal was then temporally detrended~\cite{tarvainen2002} to normalize the illumination and eliminate respiratory-induced artifacts.

\subsubsection{Correlation priors}
A transformation $T$ was sought to map the set of absorbance frames $a(x,y,t)$ to a pulsatility strength map $C(x,y)$ describing pixelwise pulsatile components:
\begin{equation}
C(x,y) = T(a(x,y,t))
\end{equation}
Rather than estimating the pulsatility based on heuristic information, which is participant-independent and may introduce uncontrolled sources of noise, this model can be augmented by incorporating prior information directly into the transformation:
\begin{equation}
C(x,y) = T(a(x,y,t) \mid z)
\end{equation}
where $z$ is the ground-truth blood pulse waveform. The ground-truth blood pulse waveform was measured synchronously with the frames, and Pearson's linear correlation coefficient was computed between the ground-truth signal and each pixel's temporal signal:
\begin{equation}
T(a(x,y,t) \mid z) = \frac{\sigma_{a z}}{\sigma_{a}\sigma_{z}} \cdot \mathbbm{1_{\mathbb{R}^+}}{\left( \frac{\sigma_{a z}}{\sigma_{a}\sigma_{z}} \right)}
\end{equation}
where $\sigma_{a z}$ is the covariance between the pixel and ground-truth signals, $\sigma_{a},\sigma_z$ are the standard deviations of the pixel and ground-truth signal respectively, and $\mathbbm{1_{\mathbb{R}^+}}$ is the indicator function of positive real numbers. This computation is scale- and offset-independent, suitable for capturing the ratio nature of the blood pulse waveform. Values of $T<0$ were set to 0 for weight computation.

\subsubsection{Importance-weighted kernel density estimation}
To quantify the spatial pulsatile strength, a physiologically-derived importance-weighted scheme was developed. An anatomical location that exhibited strong pulsatility contributed a larger weight than locations that those that contained weak or no pulsing. This system allows for continuous kernel-based probabilistic pulsatility density estimation later. The importance map for participant $i$ was computed as:
\begin{equation}
V_i(x,y) = \max \{C_i(x,y), 0\}
\end{equation}

To infer pulsatility patterns across the whole sample, the primary anatomical locations must be co-aligned. The camera was systematically and consistently setup for all participants, however, differences in anatomy, minor rotation (relative to the camera) and translation (relative to the frame region) were observed. To correct for these relative distortions, the problem was posed as a coordinate mapping problem, where each participant's weight data $V_i(x,y)$ are projected into the co-aligned pulsatility space $V(x',y')$. Mathematically:
\begin{equation}
\begin{bmatrix}
x' \\ y'
\end{bmatrix} = H \left( 
\begin{bmatrix}
x \\ y
\end{bmatrix}
\right)
\label{eq:H}
\end{equation}
where $H_i$ is a coordinate mapping function that maps $(x,y)$ from participant space $V_i$ to $(x',y')$ in the co-aligned space $V$. Note that this transformation projects points directly into a Cartesian space. There is no need for interpolation, which may have caused local inaccuracies. Implementation details of $H$ are discussed later. This aggregate co-aligned pulsatility space, $V(x',y')$, was populated with weighted points from each participant:
\begin{equation}
V(x',y') = \text{median}\{V_i(x_i,y_i)\}
\end{equation}
where $(x_i,y_i)$ are the coordinates in the participant space that project to $(x',y')$ in the aggregate space according to the mapping function $H_i$.

\begin{figure}
\centering
\includegraphics[width=\textwidth]{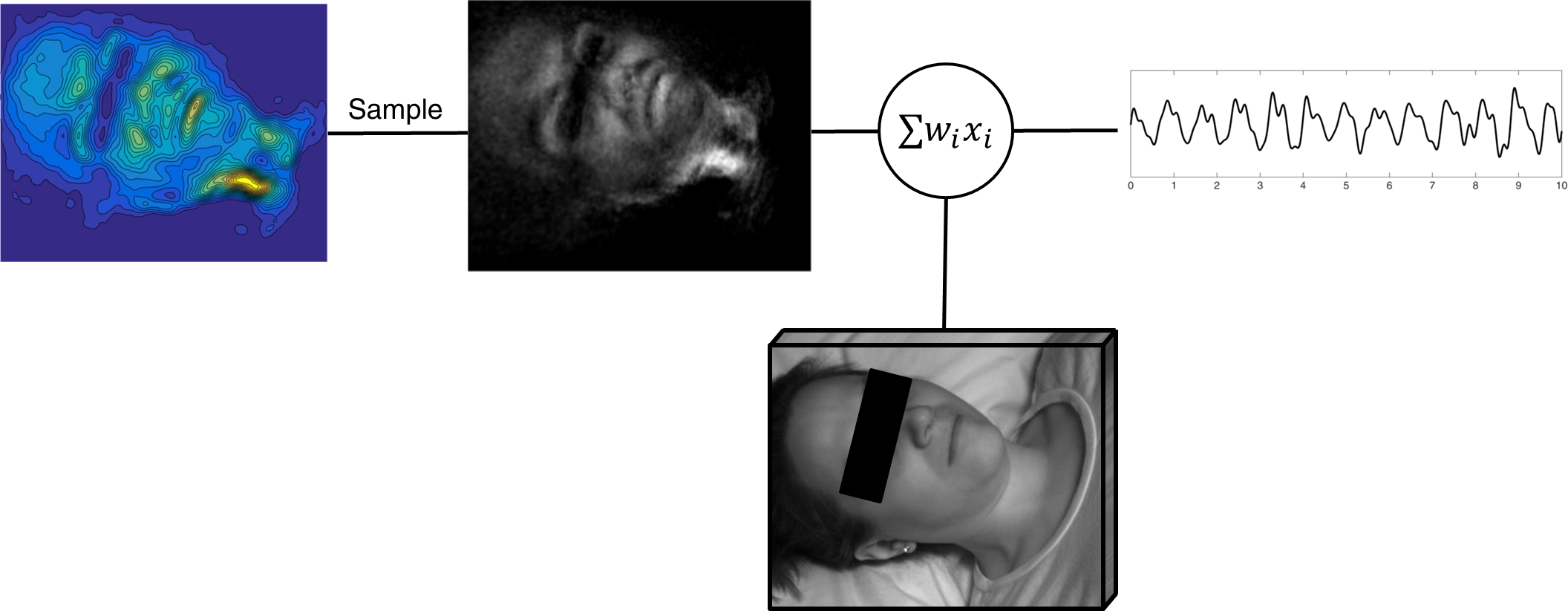}
\caption{Extracting a blood pulse waveform using the probabilistic pulsatility model. The continuous model was discretely sampled to match the resolution of the target frames, and transformed to align the match the anatomical characteristics of the participant. Pixelwise weighted averaging results in a robust blood pulse waveform.}
\label{fig:sample}
\end{figure}

A resolution-agnostic model can be computed by estimating a continuous probability density function, and sampling this density function according to the given system's resolution. A modified Parzen-Rosenblatt kernel density estimation method~\cite{rosenblatt1956,parzen1962} was used to estimate the continuous pulsatility probability density function:
\begin{equation}
p(x,y) = \frac{1}{|V|} \sum_{k=1}^{n} \frac{V(x_k,y_k)}{w^2} \Phi \left( \frac{\vec{v}-\vec{v}_k}{w} \right)
\end{equation}
where $n$ is the total number of points, $w$ is the spatial window width, $\vec{v}=(x,y)$, $\Phi$ is the window kernel, and $|V|$ is a normalization term such that $\int_x \int_y p(x,y) dy dx= 0$. The kernel is scaled according to the datum's pulsatility weight $V(x_k,y_k)$. Using the 2-D Gaussian kernel:
\begin{equation}
\Phi(\vec{u}) = \frac{1}{2\pi} \exp \left( -\frac{\vec{u}^T \vec{u}}{2} \right)
\end{equation}
the final probabilistic pulsatility model formulation becomes:
\begin{equation}
p(x,y) = \frac{1}{|V|} \sum_{k=1}^n \frac{V(x_k,y_k)}{(w \sqrt{2\pi})^2} \exp \left( -\frac{1}{2} \frac{||\vec{v}-\vec{v}_k||}{w^2} \right)
\label{eq:pmodel}
\end{equation}
where $w$ is modeled as the spatial standard deviation. This pulsatility model can be used by systems of any resolution through appropriate discrete sampling:
\begin{equation}
\Omega(x,y) = \sum_{n=-\infty}^{\infty} \sum_{m=-\infty}^{\infty} \delta(x-n \tau_x,y-m \tau_y) p(x,y)
\end{equation}
where $\delta$ is the 2-D Dirac delta function and $\tau_x,\tau_y$ are the resolution periods in the coordinate space of $p(x,y)$. A physiologically derived blood pulse waveform can be extracted using the discretely sampled pulsatility map:
\begin{equation}
z(t) = \sum_x \sum_y a_i(x,y,t) \Omega(x,y)
\end{equation}
Figure~\ref{fig:sample} shows a graphical depiction of this process.

\begin{figure}
\centering
\includegraphics[width=\textwidth]{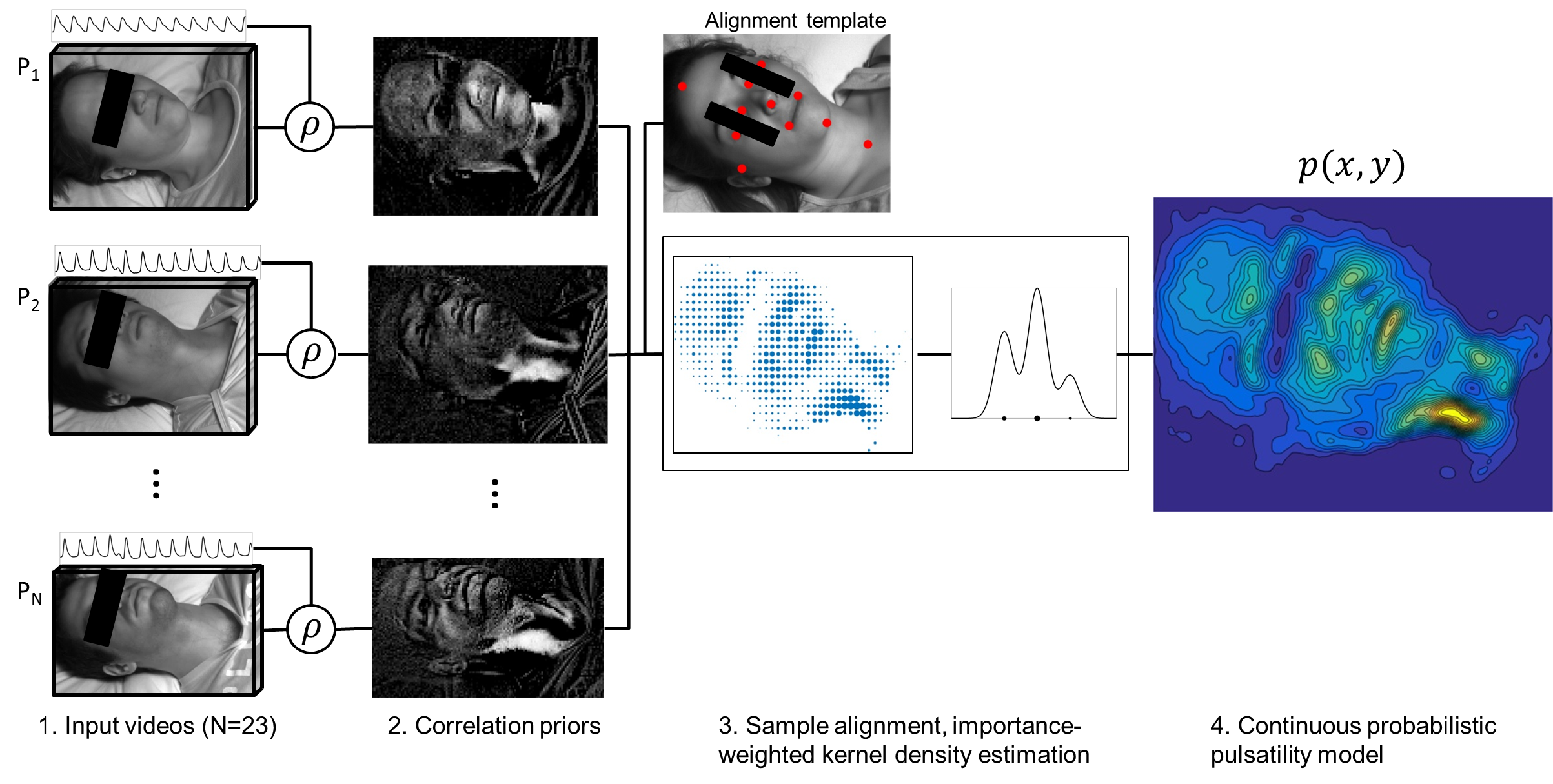}
\caption{Processing pipeline to derive the continuous probabilistic pulsatility model. Videos of 23~participants were recorded using a photoplethysmographic imaging system synchronously with the ground-truth waveform~(1). Correlation priors were computed by comparing pixelwise temporal signals against the ground-truth signal using Pearson's linear correlation coefficient ($\rho$)~(2). The correlation prior maps were aligned relative to a template and projected into a 2-D space, and spatially aggregated~(3). A 2-D kernel density estimation method was used in this space~(4) to generate the data-driven continuous probabilistic pulsatility model~(5).}
\label{fig:pipeline}
\end{figure}

\subsection{Implementation Details}
The model was built using 10~s segments for each participant. To reduce minute inter-participant spatial pulsatility differences, each frame was downsampled such that each pixel represented a $3\text{ mm}\times 3\text{ mm}$ area. We empirically found that $\sigma=3$~mm worked well for Equation~(\ref{eq:pmodel}).

We implemented the mapping function $H_i$ as a linear projective transformation~\cite{hartley2003}:
\begin{equation}
\begin{bmatrix}
x' \\ y' \\ 1
\end{bmatrix}
= \underbrace{
\begin{bmatrix}
h_{11} & h_{12} & h_{13} \\
h_{21} & h_{32} & h_{33} \\
h_{31} & h_{32} & h_{33} \\
\end{bmatrix}}_{H_i}
\begin{bmatrix}
x \\ y \\ 1
\end{bmatrix}
\end{equation}
To solve matrix $H_i$, a least-squares optimization was applied to fit fiducial markers selected on a set of frames to those same anatomical points on a template participant frame (eyes, nose, lips, chin, top and side of head, and suprasternal notch; see Figure~\ref{fig:pipeline}). Specifically, the matrix $H_i$ was solved via a least squares solution of the following linear system of equations:
\begin{equation}
\begin{cases}
x' &= h_{11}x + h_{12}y + h_{13} \\
y' &= h_{21}x + h_{22}y + h_{23} \\
1 &= h_{31}x + h_{32}y + h_{33} \\
\end{cases}
\end{equation}

\section{Results}
\subsection{Setup}
The signals extracted using the proposed probabilistic pulsatility model were compared against those extracted using the FaceMean method used in existing PPGI studies~\cite{poh2010,sun2011,sun2012,li2014}. Briefly, the pixels within the facial region found using the Viola-Jones face detection method~\cite{viola2004} were spatially averaged for each frame and concatenated, yielding a 1-D temporal signal. A leave-one-out cross-validation scheme was implemented for extracting individual participant signals. That is, participant $i$ was processed using the pulsatility density learned with the data from participants $P \setminus p_i$, where $P$ is the set of all participants and $\setminus$ is the set difference operator. The signals were temporally filtered using an ideal bandpass filter with bandwidth $[0.5,5]$~Hz (30--300~bpm).

Temporal signal fidelity was evaluated by computing the maximum cross-correlation between the extracted signal $\hat{z}$ and the ground-truth signal $z$ from the finger photoplethysmography cuff:
\begin{equation}
  \rho(\hat{z},z) = \max_{\Delta t} \left\{ \frac{\sigma_{z_t \hat{z}_{t+\Delta t}}}{\sigma_{z_t} \sigma_{\hat{z}_{t+\Delta t}}} \right\}
\end{equation}
where $\sigma_{z_t \hat{z}_{t+\Delta t}}$ is the covariance between the true and (shifted) extracted signal, and $\sigma_{z_t},\sigma_{\hat{z}_{t+\Delta t}}$ are the standard deviations of the true and (shifted) extracted signal respectively. Cross-correlation was used to account for pulsatility timing differences between the finger and face ($\Delta t \ge 0$).

Spectral signal fidelity was evaluated by computing the spectral signal-to-noise ratio (SNR) of the extracted signal:
\begin{equation}
\text{SNR}(\hat{z}) = 10 \log_{10} \left( \frac{\sum_f (Z(f))^2}{\sum_f (Z(f)-\hat{Z}(f))^2} \right)
\end{equation}
where $Z,\hat{Z}$ are the zero-DC normalized frequency magnitudes of the true and extracted signal respectively, and $f$ represents frequency. The Wilcoxon signed rank test~\cite{wilcoxon} was used to statistically compare the non-normally distributed pairwise difference between signals extracted using the proposed and FaceMean methods. Heart rate was estimated by the maximum frequency response of a modified spectral power density to reduce frequency discretization error:
\begin{equation}
\text{HR}_i = \argmax_{f_k} \sum_k Z(f_k) + Z(f_{k+1})
\end{equation}

\subsection{Data Analysis}
Table~\ref{tab:demographics} provides a summary of the sample demographics measured using bioelectrical impedance analysis. The sample contained a wide range of ages (11--60~years), body compositions (BMI 16.4--35.1~kg$\cdot$m$^{-2}$), and fair gender representation (11/13 female/male).

\begin{figure}
\centering
\includegraphics[width=\textwidth]{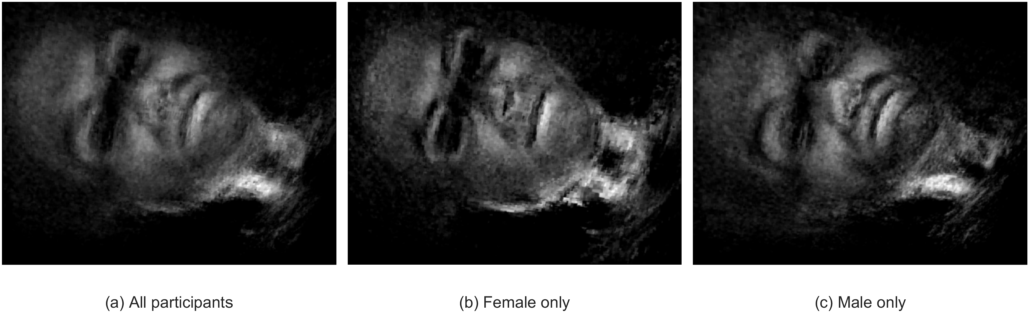}
\caption{Discretely sampled probabilistic pulsatility model across all participants (a), as well as split by gender (b,c). The effect of facial hair on reduced observed pulsatility is apparent around the mouth and chin of the male distribution (c). Both genders exhibited strong pulsatility in anatomically relevant locations, such as the neck (carotid artery) and cheeks (facial arteries).}
\label{fig:faces}
\end{figure}

Visualizing the final probabilistic pulsatility model elucidated consistent arterial pathways. Figure~\ref{fig:faces} shows the (sampled) density across all participants, as well as split by gender. The faces appeared structurally similar to a typical human face, indicating accurate sample alignment based on anatomical anchor points. The forehead, eyebrows, eyes, nose, nostrils, lips, chin and neck were all visually distinguishable. Characteristic differences existed between the female and male probabilistic faces, however the primary anatomy was consistent, leading to a cohesive combined pulsatility map. Facial hair in the male data was observed as reduced pulsatility probability in those areas due to the occlusion.

The strongest observed arterial pathway traveled across both sides of the neck. Its anatomical location was consistent with the common carotid artery, which is the primary arterial pathway to the brain and face. The common carotid artery bifurcates into the external (facial) and internal (cerebral) carotid arteries at the top of the neck. The continued blood flow traveling through the external carotid artery across the jawline could be observed in female pulsatility map (Figure~\ref{fig:faces}(b)). This phenomenon was not present in the male data (Figure~\ref{fig:faces}(c)), due to facial hair occlusion in some participants. Areas around mouth and chin exhibited reduced pulsatility compared to the female data due to facial hair occlusion. Common areas of with high pulsing probability across all participants in both genders included the carotid artery pathways, the lips, malar regions, nose, and glabella regions. Pulsing was additionally observed in females along the jaw line.

\begin{figure}
\centering
\includegraphics[width=0.7\textwidth]{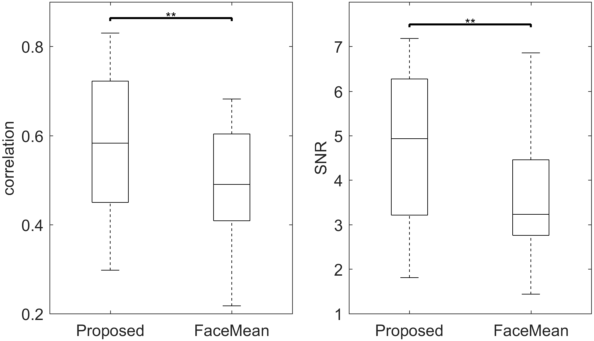}
\caption{Comparison of signal fidelity using the proposed probabilistic pulsatility model versus spatial averaging. Signals extracted using the pulsatility model exhibited statistically significantly higher pairwise correlation to the ground-truth waveform ($W=39, p<0.01$) and spectral SNR ($W=31,p<0.01$). ($^{**}\text{Wilcoxon signed rank test}, p<0.01$)}
\label{fig:boxplot}
\end{figure}

\begin{figure}
\centering
\includegraphics[width=\textwidth]{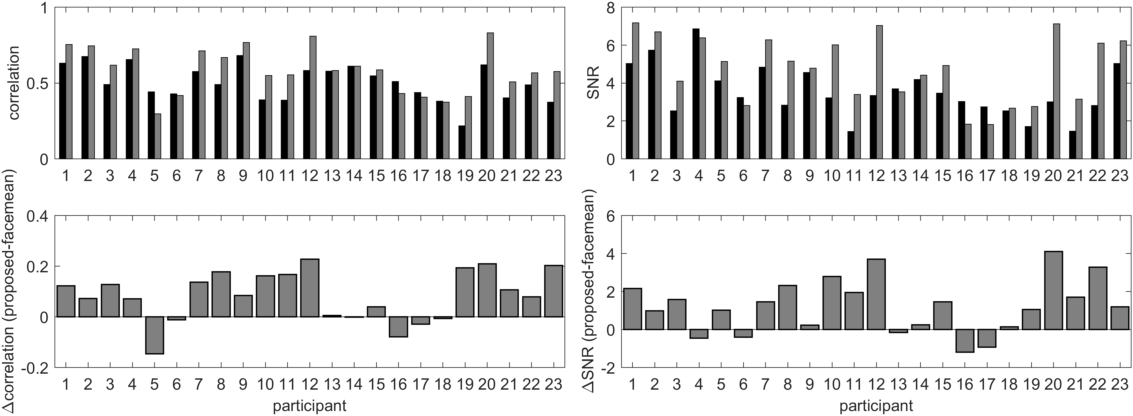}
\caption{Pairwise comparison of correlation and SNR between signals extracted using the proposed pulsatility model (black) and FaceMean (gray). Improvements were observed across most participants when using the probabilistic pulsatility model for extracting blood pulse waveform signals.}
\label{fig:barplots}
\end{figure}

Blood pulse waveform signals extracted using the probabilistic pulsatility model in a leave-one-out cross-validation scheme exhibited stronger temporal and spectral characteristics when compared to those extracted with the FaceMean method. Figure~\ref{fig:boxplot} shows the distributions of temporal correlation and SNR results across the participant sample. Combined, these measures provide signal fidelity information in both the time and frequency domain. Pairwise comparison of the signals extracted using the proposed and FaceMean methods showed that the proposed method yielded statistically significant correlation ($W=39, p<0.01$) and SNR ($W=31,p<0.01$) compared to FaceMean signals. Figure~\ref{fig:barplots} shows the pairwise differences in correlation and SNR for each of the 23 participants. Temporal and spectral fidelity improvements were observed in a large number of participants. These results indicate that the probabilistic pulsatility model identified areas of strong pulsatility

\begin{figure}
\centering
\includegraphics[width=0.8\textwidth]{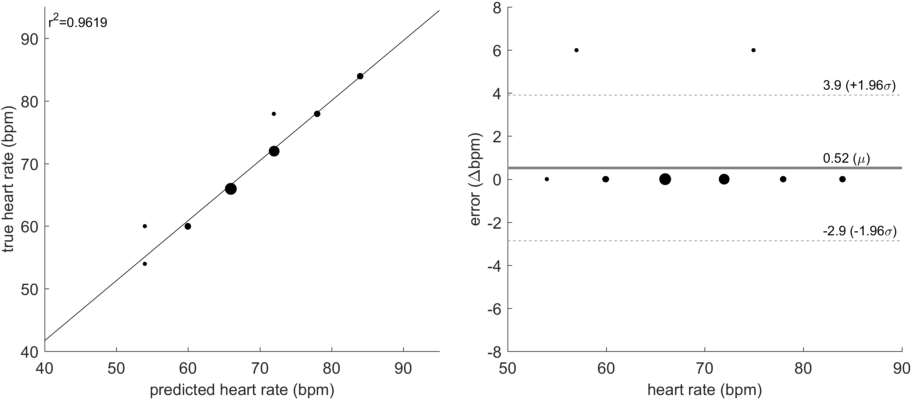}
\caption{Heart rate estimation using the pulsatility density function attained strong correlation to the true heart rate ($r^2=0.9619$) and high degree of agreement (error $\mu=0.52$~bpm, $\sigma=1.69$~bpm). Marker size indicates the number of points at that coordinate.}
\label{fig:blandaltman}
\end{figure}

Many PPGI studies investigate heart rate extraction algorithms for remote monitoring. To validate the proposed system's ability to extract heart rate, correlation and agreement to the ground-truth heart rate was investigated. Figure~\ref{fig:blandaltman} shows the Bland-Altman plot~\cite{blandaltman} for this heart rate comparison. The proposed system not only attained high correlation to the true heart rate ($r^2=0.9619$), but it also achieved strong and tight agreement (error $\mu=0.52\text{ bpm},\sigma=1.7\text{ bpm}$). This is consistent with the strong signal fidelity results in Figure~\ref{fig:boxplot}, which enable strong inferential properties of summary statistics such as heart rate.

\section{Discussion}
The results demonstrated that using a spatial probabilistic pulsatility model can enhance blood pulse waveform extraction in photoplethysmographic imaging data. The data-driven pulsatility model, which was constructed using data from a 23~participant sample with widely varying demographics, provides \textit{a priori} anatomical guidance based on consistently observed pulsatile regions.

The data-driven model is built using training data containing spatial pulsatility information. Thus, the new ``test'' data must sufficiently match the training data's spatial perspective for proper alignment. Though advanced warping models can adjust for some inconsistencies, large deviations in rotation and perspective relative to the training data may produce erroneous alignment. To address this challenge, the model was designed such that its methodology is agnostic to the type of data with which it is trained. Thus, independent training data sets can be used ($f(x,y,t)$ in Equation~\ref{eq:a}) to build custom pulsatility models suitable for the study's specific test environment. For example, some systems may find that training based on rotated viewpoints or a certain class of demographic (e.g., gender or age) may yield increased results for their specific test environment. Additionally, this model can be trained on anatomical locations other than the head, enabling whole-body cardiovascular monitoring. The model use can be extended to detect abnormal perfusion patterns that may be early markers for disorders such as peripheral vascular disease or arteriosclerosis.

The model was designed to be a continuous model so that it can be sampled by systems of any resolution. To be used, all that is required is a coordinate mapping function $H$ (from Equation~(\ref{eq:H})) for spatial alignment to the template model. In offline systems, this can be accomplished by manual or semi-automatic spatial alignment through methods~\cite{szeliski2006}. In real-time systems, automatic alignment is required, and can be accomplished using methods such as automatic face fitting~\cite{asthana2013}. In low-motion scenarios (e.g., sleeping studies, controlled experiments), a single alignment operation may be sufficient. In scenarios with increased motion, one strategy could be to calibrate the first frame to the template, and align all subsequent frames to the calibrated source frame, as it would have more similarities to frames within the same video than the model template.

\section{Conclusions}
Here, we have developed a continuous probabilistic pulsatility model that describes anatomical locations that consistently exhibited arterial pulsing across a 23~participant sample. This model can be used as prior information to enhance signal extraction in photoplethysmographic imaging systems. Since the model is a continuous model, it can be used by systems of any resolution via appropriate spatial sampling. Results showed that signals extracted using the pulsatility model exhibited statistically significant higher correlation and SNR versus unguided whole-area uniform averaging. Discretely sampled maps identified areas of consistently strong pulsing across the training data in the head, specifically the neck, malar regions, glabella regions, lips and nose. Discussions demonstrated how the model may be used in custom environments for enhanced blood pulse waveform extraction.

\acknowledgments
This work was supported by the Natural Sciences and Engineering Research Council (NSERC) of Canada, AGE-WELL NCE Inc., and the Canada Research Chairs program.


\bibliography{main}   
\bibliographystyle{spiejour}


\listoffigures
\listoftables
\end{spacing}

\end{document}